\documentclass[10pt, oneside]{article}   	
\usepackage{geometry}                		
\usepackage{graphicx}				
\usepackage{amssymb,hyperref,amsmath,multicol,color}

\title{Solar neutrino physics 
on the beginning of 2017}
\author{Francesco Vissani,\\[0.4ex]
\small INFN, Laboratori Nazionali del Gran Sasso 
{\footnotesize\em  and} 
\small Gran Sasso Science Institute, L'Aquila, Italy}
\date{}							

\begin{document}
\maketitle

\begin{abstract}
This writeup is a review of current hot topics on solar neutrinos.
It is based on a talk at the conference {\em Neutrinos: the quest for a new physics scale,} 
held at the CERN on March 2017, where the Organizers entrusted me with a 
discussion of the provocative question ``whether solar neutrino physics is over''. 
Rather than providing a straight (negative) answer, 
in view of an audience consisting mostly of colleagues working in theoretical particle physics,  
I deemed it more useful providing a description of what is the current activity of the physicists working in solar neutrinos,  
leaving the listener free of forming his/her own opinion apropos. 
\end{abstract}

\section{Introduction}

The study of solar neutrinos rests on solid scientific foundations. 
The single most important book in this field is still the one of Bahcall
\cite{Bahcall:1989ks}, a pioneer  and a renowned astrophysicist, 
but recall that  Bahcall initially was a nuclear physicist. 
His book emphasizes, already from the title, the centrality of astrophysical considerations.
 Another equally influential book is the one of Raffelt \cite{Raffelt:1996wa}, whose scientific attitude,  also in this case declared from the title, 
is the one of particle physicists instead.  The content and the points of view of these books 
are still valid and useful for the understanding of solar neutrinos. 

The only important change since the appearance  of these books 
is the proof of neutrino `oscillations', or more precisely of neutrino transformations, 
as observed by various solar neutrino telescopes and other experiments. 
Two years ago the Nobel prize committee certified the occurrence  of this phenomenon  by awarding the leaders of Super-Kamiokande and SNO experiments.
Neutrino transformations are an accepted fact and the long discussion that went under the heading of `solar neutrino anomaly' can be considered closed. Now we can proceed to compare accurately observations and expectations on solar neutrinos, testing our understanding of astrophysics and learning more interesting things. 

One year ago,  an entire issue of
Eur.~Phys.~J.~A  was devoted  to solar neutrinos~\cite{epja}; there, several updated works and useful review papers can be found.  

\section{The terms of the discussion}
There is a great interest toward the particle physics aspects concerning the neutrinos. However,  in order not to loose the correct perspective or even to undermine the chances of doing good science, it is 
essential to keep in mind that solar neutrinos 
are a typical branch of astroparticle physics.
Their  proper understanding  involves 
many fields of science, not only theoretical particle physics but also 
nuclear physics, astrophysics, astronomy and also experimental nuclear/particle physics. 
As in any field, there are some jargonic terms of the solar neutrino field cherished by the tradition and often used, 
that however make less easy to appreciate the current debate.
Therefore, we begin simply by discussing the  acronyms / main terms / keywords that recur in the discussion, and in particular: 
\textsf{SSM, ES, PP, CNO~/ helioseismology, metallicity, Boron neutrinos, Borexino.}

\begin{table}[t]
\begin{minipage}{0.7\textwidth}
{\small
\begin{tabular}{|c||l|c|c|c|}
\hline
Name & Reaction & $Q$-value [keV] & $E_\nu^{\mbox{\tiny max}}$ [keV] \\ \hline\hline
PP I & $\mbox{p}+\mbox{p}\to \mbox{D}+\beta^++\nu_e$ & 
 1442 & 420  \\ 
 & $\mbox{p}+\mbox{D} \to {}^3\mbox{He}+\gamma$ & 
 5494 &  -  \\ 
 & $ {}^3\mbox{He}+ {}^3\mbox{He}\to \alpha+2 \mbox{p} $ & 
 12860 &  \\ \hline
pep & $\mbox{p}+\mbox{p}+e\to \mbox{D}+\nu_e$ & 
 1442 & 1442  \\ \hline
PP II & ${}^3\mbox{He}+{}^4\mbox{He}\to {}^7\mbox{Be}+\gamma$ & 
 1586 &  \\ 
 & ${}^7\mbox{Be}+e\to {}^7\mbox{Li}+\nu_e$ & 
 862, 384 & 862, 384  \\ 
 & ${}^7\mbox{Li}+p\to 2\alpha$ & 
17347 &  \\ \hline
PP III & ${}^7\mbox{Be}+\mbox{p}\to {}^8\mbox{B}+\gamma $ & 
137 &  \\ 
& ${}^8\mbox{B} \to  {}^8\mbox{Be}^*+\beta^++\nu_e$ & 
$18471-E_x$ & 14600$\div$15100  \\ 
& $ {}^8\mbox{Be}^* \to 2\alpha$ & 
$E_x$ & \\ \hline
hep (PP IV) &
 ${}^3\mbox{He}+\mbox{p}\to\alpha+\beta^++\nu_e$ & 
 19795 & 18773  \\ \hline \hline
CNO-I 
& ${}^{12}\mbox{C}+\mbox{p}\to{}^{13}\mbox{N}+\gamma$ &1943 &\\ 
& ${}^{13}\mbox{N} \to{}^{13}\mbox{C}+\beta^++\nu_e$ &2221 & 1199   \\ 
& ${}^{13}\mbox{C}+\mbox{p}\to{}^{14}\mbox{N}+\gamma$ &7551 & \\ 
& ${}^{14}\mbox{N}+\mbox{p}\to{}^{15}\mbox{O}+\gamma$ & 7297 &\\ 
& ${}^{15}\mbox{O} \to{}^{15}\mbox{N}+\beta^++\nu_e$ & 2754 & 1732  \\
& ${}^{15}\mbox{N}+\mbox{p} \to{}^{12}\mbox{C}+\alpha$ & 4966 &  \\  
\hline
\end{tabular}}
\end{minipage}
\hfill
\begin{minipage}{0.27\textwidth}
\caption{\footnotesize\em Nuclear reactions in the Sun, adapted from \cite{Bahcall:1989ks}.
The first 11 reactions are the {\rm PP} chain,  grouped in 5  branches; 
the last 6 are part of cold CNO cycle that contributes little to solar luminosity.
The 2nd reaction of {\em PP II} branch is an electron capture and produces two neutrino lines; the 2nd reaction of the {\rm PP~III} branch depends on the energy of the excited {\rm ${}^8$Be} state $E_x$ that is not known precisely~\cite{Winter:2004kf}. 
Particles or atomic nuclei are indicated;  
 $\mathrm{p}={}^1\mathrm{H}$ and $\mathrm{D}={}^2\mbox{\rm H}$.
For final states the traditional Rutherford's 
notation 
$\alpha={}^4\mathrm{He}$  and 
$\beta^+=\mathrm{e}^+$ is used. The energy of the positron is included in $Q$.
\label{tab:jb}}
\end{minipage}
\end{table}

\paragraph{Expectations}
The nuclear physics processes by which the Sun produces energy are listed in table~\ref{tab:jb};
they can be grouped in a chain of reaction named \textsf{PP} chain and in the catalytic 
\textsf{CNO} cycle.  
An important locution is that of `Standard Solar Model' whose acronym is \textsf{SSM}.
It is used either generically or  to denote the model originally developed 
by J.~Bahcall and improved by several collaborators. This was and it is a tool of essential 
importance for the investigations.   It is the comparison of the \textsf{SSM} and of the measurements in Homestake, Kamiokande, 
GALLEX/GNO, SAGE, that 
proved the existence of physics beyond the standard model, see e.g., \cite{Adelberger:1998qm} and 
convinced the scientific community that the 
`solar neutrino anomaly' was worth investigating further. 
Even the formulation of the principle of the SNO detector
\cite{Chen:1985na} relied heavily on \textsf{SSM}, despite of the fact that the 
measurement of the neutral current events of SNO is often termed a `model independent test'. 
The study of the oscillations of solar surface, 
connected to the understanding of  the matter distribution in the outermost layers of the Sun, a discipline called \textsf{helioseismology}, 
has lead to overall and independent validation of the \textsf{SSM}.
Of particular importance are the trace elements beyond hydrogen and helium, 
called collectively  `heavier elements' or referred as 
\textsf{metals} in astrophysical parlance.
The predicted inner structure of the Sun  depends upon their abundance.
The existing and supposedly precise measurements   
leads however to contradicting \textsf{SSM}  predictions~\cite{Bahcall:2004pz}, a problem that to date is still unsolved.
The most recent version of the \textsf{SSM}
appeared just a few months ago   \cite{Vinyoles:2016djt}; 
this study assesses the missing information,  quantifies the shortcomings of the model and furthermore examines the possible key tests.

 {\footnotesize
\begin{table}[t]
\begin{center}
\begin{tabular}{ccccc}
Experiment   & Experiment  & Neutrino & Energy& Minimum \\ 
Name             & type  & detection reaction & meas. & $\nu$ energy\\ \hline
Homestake  & {\small radiochemical} & $\nu_e +{}^{37}\mbox{Cl}\to e+{}^{37}\mbox{Ar}$\ \ \ {\footnotesize (CC)} & no & 814 keV \\
Gallex/GNO  & {\small radiochemical}  & $\nu_e +{}^{71}\mbox{Ga}\to e+{}^{71}\mbox{Ge}\:$ {\footnotesize (CC)} & no & 233 keV \\
SAGE   & {\small radiochemical} & $\nu_e +{}^{71}\mbox{Ga}\to e+{}^{71}\mbox{Ge}\:$ {\footnotesize (CC)} & no &233 keV \\
SNO   & {\small heavy water}  & $\nu +D\to \nu+p+n$\ \ \ \ \ \  {\footnotesize (NC)} & no &$\!\!$2230 keV \\
SNO   &    {\small heavy water} &   $\nu_e +D\to p+p+ e $ \ \ \  \  {\footnotesize (CC)} & yes &$\!\!$4940 keV \\
Kamiokande  & {\small water Cherenkov} & $\nu +e\to \nu+e$\ \ \ \ \ \ {\footnotesize (CC+NC)} & yes &$\!\!$7750 keV \\
Super-K   & {\small water Cherenkov} & $\nu +e\to \nu+e$\ \ \ \ \ \  {\footnotesize (CC+NC)} & yes &$\!\!$3730 keV\\
Borexino   & {\small ultra-pure scintillator} & $\nu +e\to \nu+e$\ \ \ \ \ \  {\footnotesize (CC+NC)} &yes &  285 keV\\
\end{tabular}
\end{center}
\caption{\footnotesize\em Main characteristics of solar neutrino experiments, including the  the minimum neutrino energy that yields an observable signal.
The first 4 entries
refer to experiments that {\bf count} the nuclei or neutrons due to neutrino interactions.
The last 4 entries instead 
describe experiments capable to {\bf measure the energy} of the electrons in the final state with 
energies above 3.5, 7.5, 3.49 and 0.15 MeV respectively.
\label{tab:solarExp1}}
\end{table}}

\paragraph{Experiments} 
It is funny to note that there are just two solar neutrino experiments that have been realized mostly 
to study the functioning of the Sun: the first one,  Homestake and the last one, \textsf{Borexino}.
In the meantime we had other excellent experiments, which contributed to the field but which were aimed, mostly, to understand neutrinos rather than the Sun--as summarized in Table~\ref{tab:solarExp1}.
There are many differences among these experiments. Homestake and the Gallium experiments observe all neutrino interactions above a certain energy threshold, 
using neutrino interactions on nuclei, whose  cross sections are known less precisely than those used by the other experiments. 
Kamiokande and its successor Super-Kamiokande (often denoted as Super-K), 
SNO and \textsf{Borexino}, instead, have been able to observe the differential spectrum. 
Let us recall that \textsf{Borexino}, just as Kamiokande/Super-Kamiokande,
detects solar neutrinos thanks to the {neutrino elastic scattering on atomic electrons}, \textsf{ES} for short, namely, 
$$
\nu+e\to \nu +e 
$$
The cross section of \textsf{ES} is theoretically very clean; radiative corrections in the standard model 
are known and the precision is much better than what currently needed. 
However, in scintillators detectors such as \textsf{Borexino}, KamLAND or the future detector JUNO, 
the direction of the single electron cannot be reconstructed and 
background cannot be discriminated: for this reason, it is necessary to be sure {\em a priori} 
of the absence of radioactive contaminants. 
\textsf{Borexino} proved that a condition of ultrahigh \textsf{radio-purity} 
is achievable and moreover the light yield of electrons is high, which allows one to measure the energy precisely, with few~\% precision, down to very low energies. This experiment succeeded to achieve an impressive energy  
threshold of about 150 keV.\footnote{The minimum neutrino energy is when the  final state neutrino recoils backward.
Summing the  energy and momentum conservation conditions $E_\nu+m_e=E_\nu'+E_e$ and  
 $E_\nu=-E_\nu'+p_e$ we get  
$E_\nu^{\mbox{\tiny min}}=(T_e+p_e)/2$.}

\begin{figure}[t]
\begin{minipage}{0.65\textwidth}
\centerline{\includegraphics[width=0.8\linewidth]{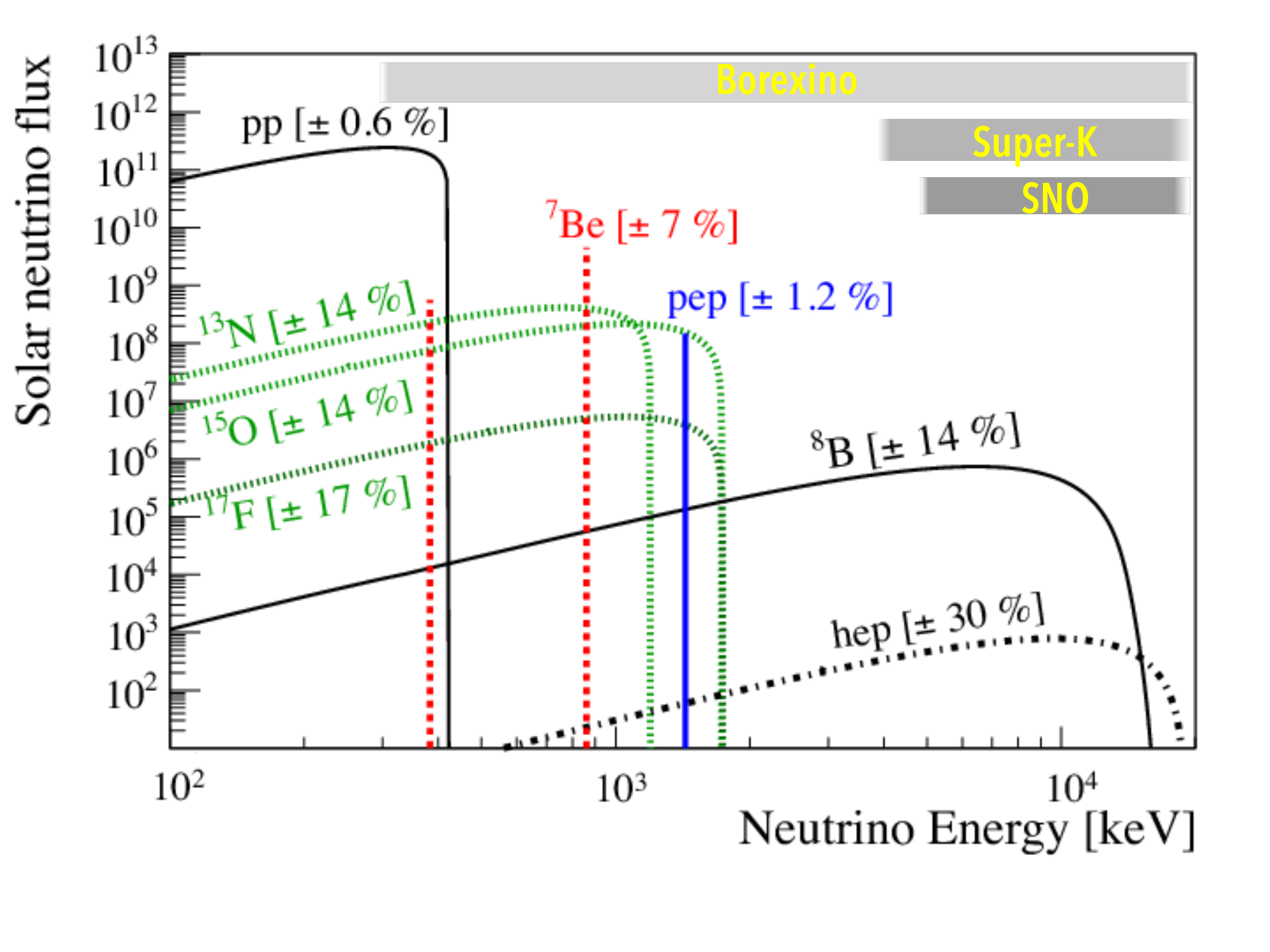}}
\end{minipage}\hskip-2mm
\begin{minipage}[c]{0.35\textwidth}
    \caption{\footnotesize\em The solar neutrino spectrum predicted by the \textsf{SSM} of Bahcall and collaborators,
    as  compiled by~\cite{DAngelo:2014lmb}. 
    The three grey strips indicate the energy ranges of operation of the solar neutrino telescopes able to measure the differential spectrum. Compare with Tables~\ref{tab:jb} and \ref{tab:solarExp1}; see the text for discussion. 
  \label{fig:fluxexp}  }
  \end{minipage}
  \vskip-8mm
  \end{figure}

\section{Nuclear astrophysics}

\paragraph{The PP chain}
In Fig.~\ref{fig:fluxexp} we summarize the expectations for solar neutrinos as a differential spectrum, 
and show the range of energies explored by  Super-K, SNO and \textsf{Borexino}, namely the solar neutrino experiments that 
are capable to measure the energy of the events and to separate the various neutrinos. 

Let us begin by discussing the high energy solar neutrinos. Those 
produced in the \textsf{PP III} branch of the \textsf{PP} chain in Table~\ref{tab:jb}, the so called \textsf{Boron neutrinos}, 
have been investigated accurately thanks to all these experiments. 
Super-K and SNO contributed mostly, since 
\textsf{Borexino}  is by far smaller. The latter, however, probed \textsf{Boron neutrinos} till the lowest energies ever achieved 
and it is planning to go down till 2.5 MeV.\footnote{The name \textsf{``Borexino''} derives from the one of a previously proposed experiment, 
BOREX, that was supposed to be loaded with ${}^{11}$B and that  was aimed at observing \textsf{Boron neutrinos} \cite{raju}. The concept of the experiment was radically modified but the connection with boron remained in the name and thanks to the results on \textsf{Boron neutrinos}.}
The highest energy neutrinos, so called \textsf{hep} belonging to \textsf{PP IV} branch, are still unobserved to date owing to the small flux; their search is continued by Super-K and will be continued in future large detectors as  
Hyper-K and maybe JUNO.

However, the above neutrinos amount only to 0.02\% of the neutrinos emitted by the Sun. 
All the other neutrinos belonging to the \textsf{PP} chain, namely, \textsf{PP, pep, Beryllium,} 
have been observed directly only by \textsf{Borexino}. A few remarks on \textsf{PP} neutrinos are in order. 
As evident from  Fig.~\ref{fig:fluxexp},  these neutrinos 
are the most abundant among solar neutrinos and those directly related to the principal nuclear chain of energy generation in the Sun. 
Thus, it is possible to say that, after the observation of \textsf{PP} neutrinos, we have sound experimental bases  to the basic  \textbf{understanding of how the Sun functions}. It should be emphasized that this measurement is 
quite recent, having been obtained just 3 years ago \cite{Bellini:2014uqa}.
Note that solar neutrinos flow out immediately after being produced, whereas the electromagnetic radiation does it several 100 thousands years later. Since this is a small time in comparison to its lifetime, it is possible (and easy) to relate the flux of \textsf{PP} neutrinos just to the solar luminosity observed today:
this makes very reliable the theoretical expectation. 
Conversely, the observation of  \textsf{PP}  neutrinos ensures us that the Sun will continue to work as it is currently doing for several 100 thousand years at least. The  \textsf{SSM} (i.e., the theory) predicts that the increased luminosity of the Sun will vaporize terrestrial oceans after one billion years; then, 
four billion years later or so, its life as a star will end.

\begin{figure}[t]
\begin{minipage}{0.62\textwidth}
\includegraphics[width=0.64\linewidth]{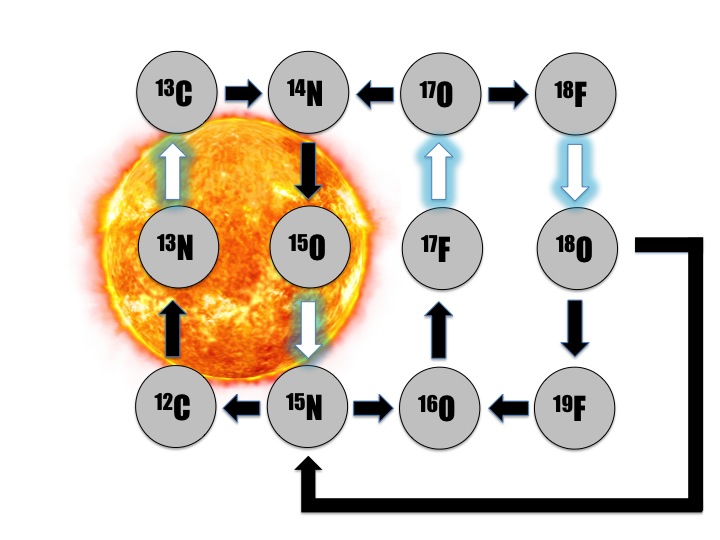}
\end{minipage}
\hskip-2cm
\begin{minipage}{0.45\textwidth}
    \caption{\footnotesize\em The cold \textsf{CNO} cycles.  The white arrows denote $\beta^+$-decays where $\nu_e$  are also released. 
    The leftmost loop is the \textsf{CNO-1} cycle relevant for the Sun.    Compare with Table~\ref{tab:jb} and Fig.~\ref{fig:fluxexp}.
  \label{fig:cno} }
  \end{minipage}
  \end{figure}

\paragraph{The CNO cycle} 
The \textsf{CNO} cycle  depicted in Fig.~\ref{fig:cno} is the main cycle of energy generation in the most massive stars. 
This is expected to yield $\sim 1$ \% of the solar luminosity and it is not explored yet. 
Its measurement may help us to fix the pending issues of \textsf{SSM} and moreover it is a unique occasion to 
check the reliability of our understanding of stellar astrophysics. As it is evident from Table~\ref{tab:jb} and Fig.~\ref{fig:fluxexp}, 
the corresponding neutrinos have low energies and the only experiment that has a chance to observe them to date 
is, once again, \textsf{Borexino}. The signal related to the \textsf{ES} reaction is illustrated in Fig.~\ref{fig:cno}. The left panel emphasizes its characteristic shape; the right panel instead shows that the signal of interest occurs in the same region of the one of pep neutrinos, that however are quite well-known being 
connected to the \textsf{PP} neutrinos and to the solar luminosity.  A few remarks are in order:\\
{\em 1)} The relative abundances of the species which contribute to the \textsf{CNO} catalysis of hydrogen in helium  
are precisely predicted by nuclear physics, since they are part of a cycle, 
whereas the absolute abundance is not and depends upon the model; in the right panel we use the model 
with high \textsf{metallicity} of Grevesse-Sauval (GS98). \\
{\em 2)} Due to the presence of ${}^7$Be neutrinos and to the finite resolution of the detector, 
it is not possible to see the \textsf{CNO} neutrinos for very low energies, $\sim$0.7 MeV in the figure.\\
{\em 3)}  \textsf{Borexino} contains 278 t of pseudocumene with brute chemical formula C$_{9}$H$_{12}$ and it is working since 10 years.\\
All in all, the signal is not small and the detection does not seem impossible.

\begin{figure}[b]
\vskip-4mm
\includegraphics[width=0.43\linewidth]{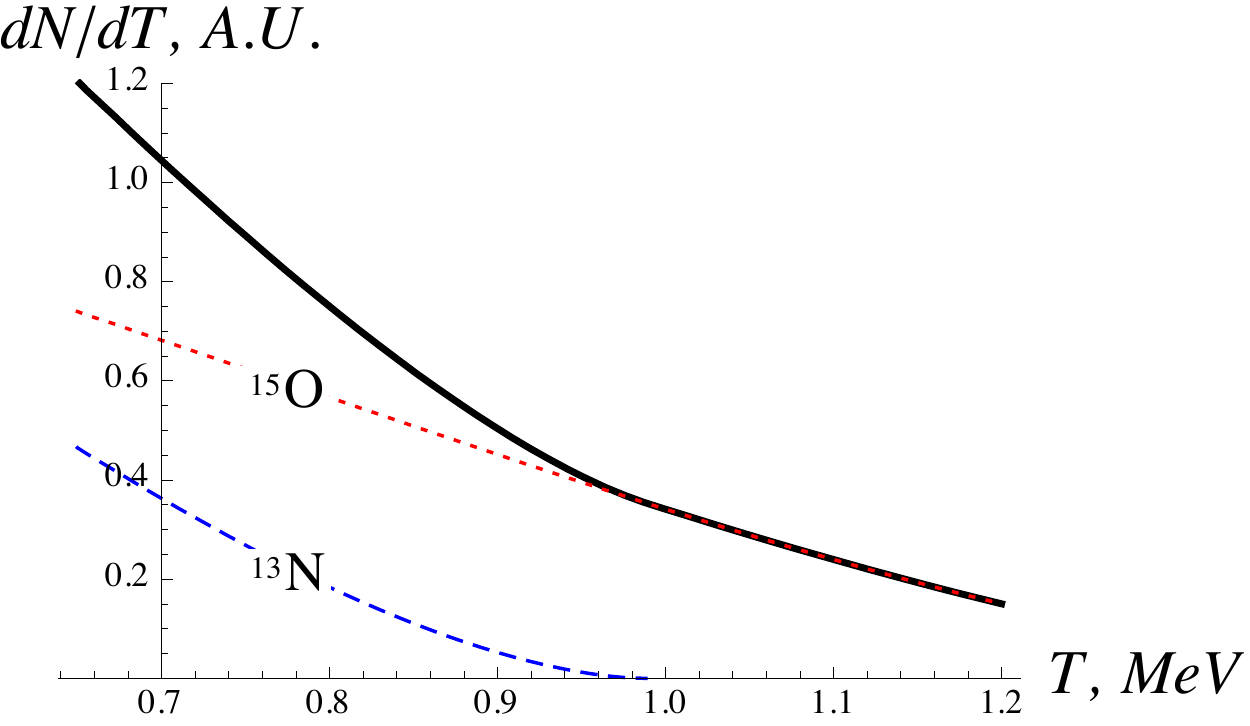} \hfill \includegraphics[width=0.42\linewidth]{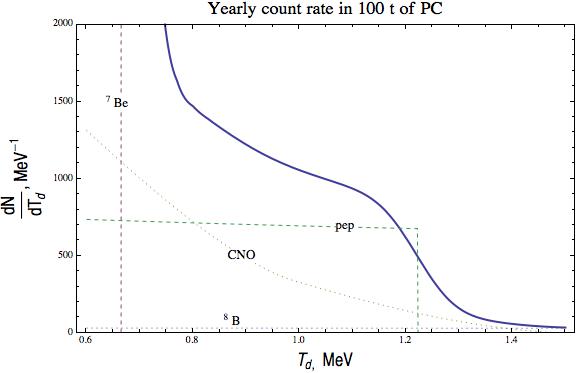}
    \caption{\footnotesize\em Spectra due to \textsf{ES} reaction due to \textsf{CNO} neutrinos as a function of the kinetic 
    energy $T_e$ of the final state electron.   
    Left panel: the two main components and the sum spectrum, with the characteristic feature at $\sim$1 MeV 
    (the ${}^{17}$F neutrinos have almost the same shape as the     ${}^{15}$O ones 
    but are much less abundant, see Fig.~\ref{fig:fluxexp}).
    Right panel: yearly count rate in 100 t of pseudocumene (PC), 
    including the effect of energy resolution of 5\% at $T_e$=1 MeV and showing also the relevant branches of 
    \textsf{PP} neutrinos. 
  \label{fig:cno} }
  \end{figure}

\begin{figure}[t]
\begin{minipage}{0.55\textwidth}
\includegraphics[width=1\linewidth]{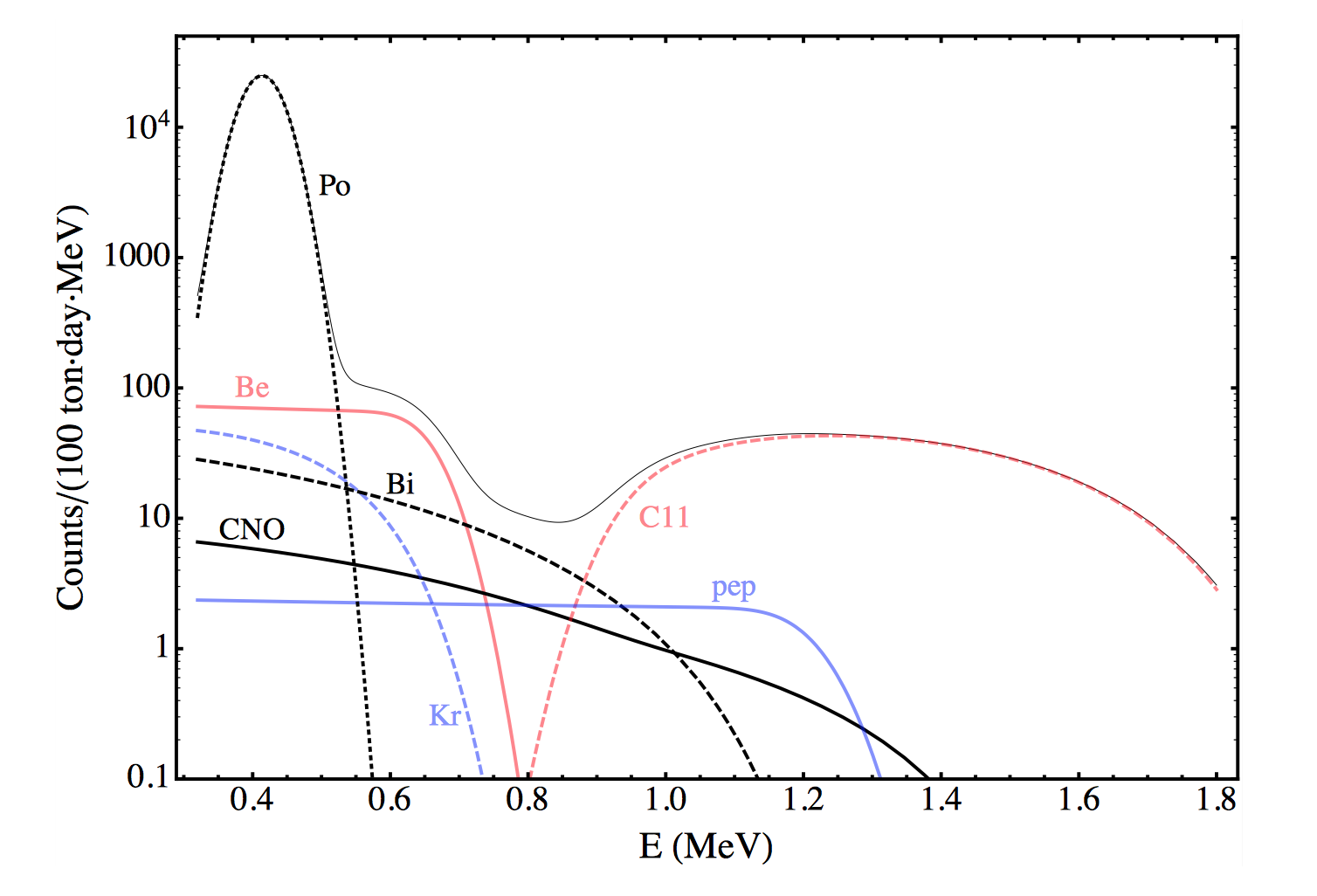}
\end{minipage}\hfill
\begin{minipage}[c]{0.4\textwidth}
    \caption{\footnotesize\em Spectra of the \textsf{ES} signal due to solar neutrinos  (continuous lines) and to background processes
    (dashed lines). This plot corresponds to the situation in 2011,  with 25~count per day/100t of ${}^{11}$C
    (before three-fold tagging) and with 20~count per day/100t of ${}^{210}$Bi. From~\cite{Villante:2011zh}.   
     \label{fig:tag} }
     \end{minipage}
  \end{figure}

The real problem is the occurrence of other (background) phenomena due to 
radioactive contaminants, that could mimic the signal due to \textsf{CNO}  neutrinos in which we are interested. 
At higher energies there is the $\beta^+$ decay of ${}^{11}$C, which can be 
lowered by more than order of magnitude implementing a three-fold coincidence  and that can be 
further tagged by discriminating the $\beta^+$ from $\beta^-$ \cite{Bellini:2013lnn}. At lower energies, which is worse,
there is instead a $\beta^-$ decay  ${}^{210}$Bi, which resembles closely the signal. These considerations are quite evident from 
Fig.~\ref{fig:tag} taken from \cite{Villante:2011zh}. 
The last reference   is a theoretical investigation  of possible strategies aimed at 
coping with the ${}^{210}$Bi background. The basic point made there is quite simple; the 
${}^{210}$Bi is part of a chain of decays, which includes 
$$
\begin{array}{rcc}
{}^{210}\mbox{Pb}
& \stackrel{\mbox{\tiny32.3 y}}{\xrightarrow{\hspace*{0.7cm}}}&
{}^{210}\mbox{Bi}+\bar\nu_e+\beta \\[1ex]
{}^{210}\mbox{Bi}
& \stackrel{\mbox{\tiny7.2 d}}{\xrightarrow{\hspace*{0.7cm}}}&
{}^{210}\mbox{Po}+\bar\nu_e+\beta \\[1ex]
{}^{210}\mbox{Po}
& \stackrel{\mbox{\tiny200 d}}{\xrightarrow{\hspace*{0.7cm}}} &
{}^{206}\mbox{Pb}+ \alpha 
\end{array}
$$
This implies a close relationship between the $\beta$ from ${}^{210}$Bi decay and the (intense) signal due to $\alpha$ particle from 
${}^{210}\mbox{Po}$ decay, visible in the leftmost part of Fig.~\ref{fig:tag}. 
This gives a chance to measure 
the ${}^{210}\mbox{Bi}$ contribution,\footnote{E.g., if the detector is stable and no 
further contaminants are introduced, 
the average \textsf{activities} $n$ (=the ratio of the number of 
nuclei over the lifetime) 
of the   ${}^{210}\mbox{Bi}$ and ${}^{210}\mbox{Po}$ species are given by $ n_{\mbox{\tiny Po}}(t)= n_{\mbox{\tiny Bi}}+  (n_{\mbox{\tiny Po}}(0)-n_{\mbox{\tiny Bi}} ) \exp(-t/\tau_{\mbox{\tiny Po}}) $;
the  condition of secular equilibrium obtains for $\tau_{\mbox{\tiny Pb}}\gg t\gg\tau_{\mbox{\tiny Po}}$.}
that can be subtracted from the  region of the \textsf{CNO} signal, 
that in this way becomes observable~\cite{Villante:2011zh}.

The reason why since September 2015 \textsf{Borexino} is wrapped in thermal insulation  (as can be seen also from Wikipedia \cite{wiki}) is to keep thermal fluctuations under control, avoiding to reintroduce  radioactive contaminants in the detector and  attempting the extraction of the \textsf{CNO} signal. In a recent PhD thesis of a member of \textsf{Borexino}, successfully defended at the GSSI on December 2016, one can read the following words~\cite{ilia},
\begin{quotation}
{\em ... Borexino detector has statistical sensitivity to \textsf{CNO} 
and \textsf{pep} 
neutrinos when the dedicated analysis here developed is applied. Central values are $(5.2\pm 1.8_{stat})\times 10^8$ cm$^{-2}$s$^{-1}$ and  $(1.31\pm 0.35_{stat})\times 10^9$ cm$^{-2}$s$^{-1}$ ...}
\end{quotation}
that, apart from the misprint ($10^9$ should be $10^8$ as in Figs.~\ref{fig:cno} or~\ref{fig:tag}),  
indicates that we are in an exciting moment; developments in the next months are expected.

\section{Particle physics}

\paragraph{The Mikheyev-Smirnov-Wolfenstein theory}
The transformation of the electron neutrinos is governed by the interplay between the 
vacuum oscillation phenomena and the fact that the electron neutrinos propagate differently from the other neutrinos in matter, owing to the fact that they are the only one to undergo charged-current interactions with the electrons of the medium as argued by Wolfenstein \cite{w}. 
The physics is very well-understood and also amenable to a simple description thanks to the analysis of Mikheyev and Smirnov \cite{ms}. 
For instance the neutrinos that reach the detector on day time are in a good approximation (where we neglect the effect of the mixing angle $\theta_{13}$) 
described by the formula,
$$
P_{ee}^{\mbox{\tiny day}}(E_\nu)\approx \frac{1}{2}\left( 1+ \cos2\theta\times \cos2\theta_m \right)
\mbox{ with }
\cos2\theta_m =\frac{\cos2\theta-\varepsilon_\odot}{\sqrt{\sin^2 2\theta + (\cos2\theta-\varepsilon_\odot)^2}}
$$
where $\theta=\theta_{12}\approx 33^\circ$ and 
$$
\varepsilon_\odot=\frac{\sqrt{2} G_F N_e^\odot}{\Delta m^2/(2 E_\nu)}\approx 1.04 
\left( \frac{N_e^\odot}{100\mbox{ mol}} \right)
\left( \frac{7.37\times 10^{-5}\mbox{ eV}^2}{\Delta m^2} \right)
\left( \frac{E_\nu}{5\mbox{ MeV}} \right)
$$
where we have used typical values of the 
electron neutrino density $N_e^\odot$ in the region where neutrinos are produced,
of the neutrino energy $E_\nu$ along with the best fit value for $\Delta m^2=\Delta m^2_{12}$ as indicated by 
current global analyses of the available data, and mostly by \textsf{KamLAND} reactor antineutrino 
experiment,  
based on the assumption that there are 3 light but massive neutrinos.

\begin{figure}[t]
\begin{minipage}{0.35\textwidth}
    \caption{\footnotesize\em  
    Illustration of the effect of lowering the value of $\Delta m^2$ for the solar neutrino survival probabilities 
    $P_{ee}$ of solar neutrinos. 
   We use the formulae given in the text with the parameters 
    $N_e^\odot=100 \mbox{\rm\ mol/cc}$ and 
    $N_e^\oplus=2 \mbox{\rm\ mol/cc}$; 
    $\Delta m^2(\mbox{\rm KamLAND})=7.37\times 10^{-5}$ eV$^2$ and 
    $\Delta m^2(\mbox{\rm Super-K})=4.9\times 10^{-5}$ eV$^2$. 
     \label{fig:mmm} }
     \end{minipage}
     \hfill
\begin{minipage}{0.5\textwidth}
\includegraphics[width=1\linewidth]{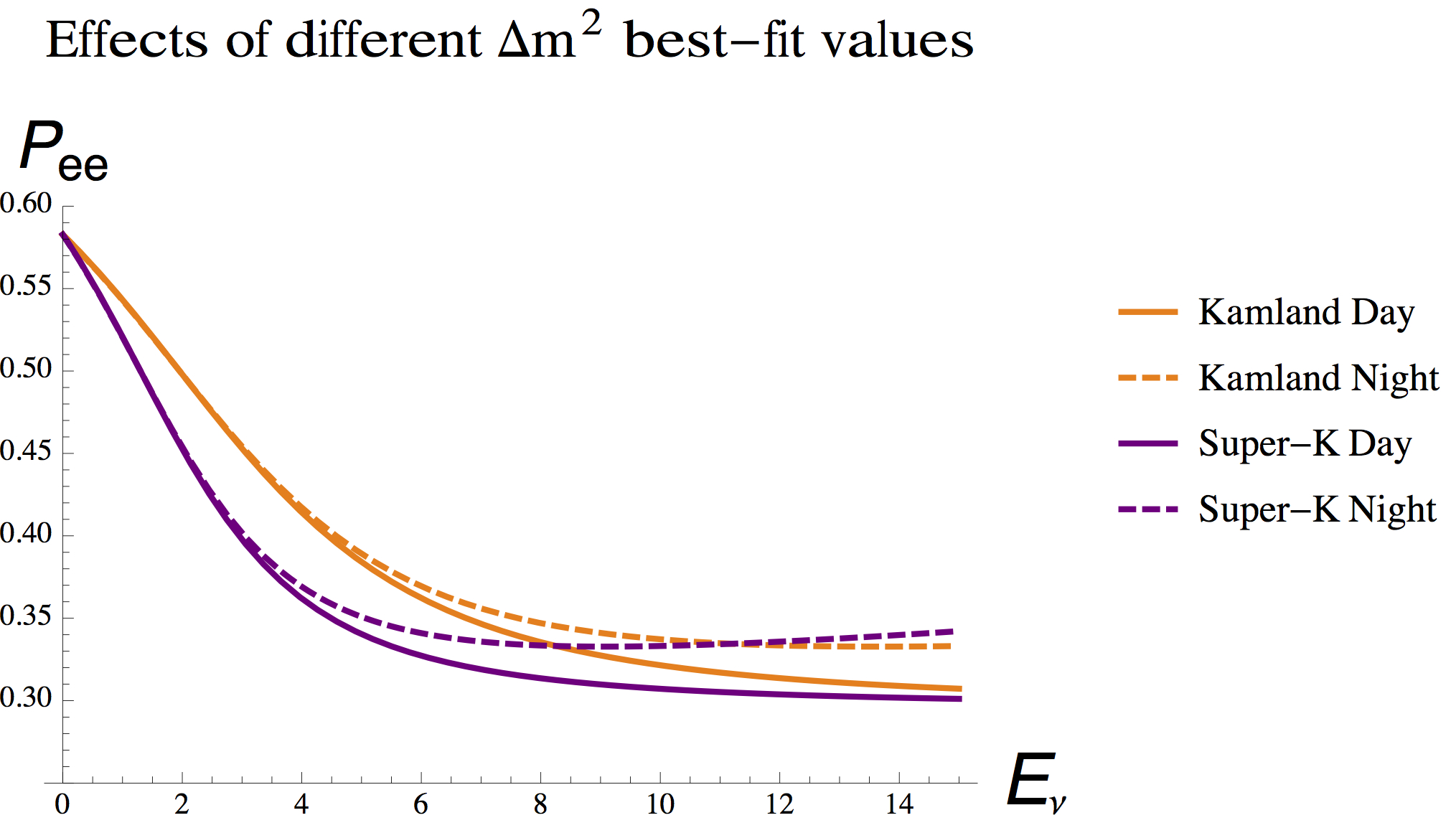}
\end{minipage}
  \end{figure}

Interestingly, electron neutrinos that arrive on night time are more abundant, thus the Sun shines {\em brighter} in neutrinos during the night! 
To understand the physics 
it is sufficient to consider the case when the neutrinos traverse
a slab of terrestrial matter of constant density. (An excellent discussion is in \cite{lisis}.)
The formula is,
$$
P_{ee}^{\mbox{\tiny night}}(E_\nu) -P_{ee}^{\mbox{\tiny day}}(E_\nu)
=\frac{\varepsilon_\oplus\times \sin^22\theta }{{\sin^2 2\theta + (\cos2\theta-\varepsilon_\oplus)^2}}\times
\frac{1/2- P_{ee}^{\mbox{\tiny day}}(E_\nu)}{\cos2\theta}
$$
Here $\varepsilon_\oplus$ has just the same  formal expression as $\varepsilon_\odot$ but the electron density in the Earth 
is of course  smaller, $N_e^\odot\gg N_e^\oplus$ and thus  $\varepsilon_\odot\gg \varepsilon_\oplus$.
To date the various observations of solar neutrinos can be accounted for in this simple setup apart from some tension  in the overall interpretation, that can be attributed to the following two facts,\\
{\em 1)} One would expect that, moving toward  the lowest  solar neutrino 
energies currently measured, the probability of oscillation should slightly rise, however
no sign of \textsf{turn up} is still perceivable. \\
{\em 2)} One would not expect that the difference between day and night is very large, however 
Super-Kamiokande measures  an effect of electron neutrino \textsf{regeneration} 
that at central value is 
twice as expected (with weak significance).\\
See \cite{skd} for a recent summary and discussion.\\
Interestingly,  both deviations could be accounted for at the same time, simply assuming that $\Delta m^2$ is smaller than stated above,  say $4.9\times 10^{-5}$ eV$^2$ as indicated by solar neutrino data alone (dropping 
\textsf{KamLAND}  results from the global fit): see Fig.~\ref{fig:mmm} for an illustration.  
Another very dramatic and almost certainly premature interpretation is in terms of CPT violation (!!!) namely one could be lead to believe  the \textsf{KamLAND} \underline{antineutrinos} and the solar \underline{neutrinos} oscillate with different parameters.
But most plausibly the two results are not 
due to the same reason and/or are just fluctuations:
 after all, \textsf{Borexino} verified that the probability of oscillations grows at low energy just as expected.

\paragraph{Speculations}
There are many other topics, of great interest for theorists working on particle physics, 
that can be investigated by means of solar neutrinos. In the following list, we cite some possible 
manifestations/new particle  with a brief comment on the related observable/phenomena,
\begin{itemize}
\item \textsf{other light (sterile) neutrinos} [shape of the spectrum; neutral current events]
\item \textsf{new oscillations on cosmic scales} [low energy data]
\item \textsf{neutrino decay} [flavor structure]
\item \textsf{neutrino magnetic moments} [solar antineutrinos]
\item \textsf{non-standard interactions} [new matter-effect]
\item \textsf{axions} [energy loss]
\item \textsf{WIMP in the Sun} [solar structure; high-energy neutrinos]
\end{itemize}
etc. This list should make evident what is 
the scientific potential of solar neutrinos for particle physics.
Some of them have stronger motivations, some less.
Since we do not have evidence of them to date, it seems fair to 
call them collectively ``speculations'',  until new relevant facts will require a change. 
We will not enter into a detailed discussion of these topics;
we will comment briefly on the first two issues, 
since they are connected to the existence of {\em other} neutrinos 
and they could affect the phenomenology of neutrino oscillation 
(just recognized by the Nobel prize committee).

First of all, it is important to stress that, even if 
it is possible to find several particle physics models that include sterile neutrinos lighter than $\sim 1$ eV, 
it is not always clear what are the theoretical motivations for this position. This is quite distinct from the question on 
whether we have some motivation from experiments, or more precisely, 
from the interpretation of some experimental fact. To date, one of the few theoretically motivated model 
is the one based on the concept of mirror symmetry. We describe here the case of exact mirror symmetry,
making reference to \cite{venya} for a complete discussion. 
In this model, there is a global $Z_2$ symmetry  
that enforces the existence of mirror bosons and fermions (quarks and leptons) including neutrinos. 
The mirror particles interact with mirror gauge bosons but do not interact directly with the ordinary ones. In this manner, 
the mirror baryon may play the role of dark matter. Moreover, mirror neutrinos are exactly degenerate in mass with ordinary neutrinos. In presence of small 
interactions, e.g., Planck scale suppressed interactions, the ordinary and mirror neutrinos with given mass mix maximally and can give rise to 
new oscillations at very long scale. This kind of effect can be tested by studying the lowest energy solar neutrinos. 

\begin{figure}[t]
\includegraphics[width=1\linewidth]{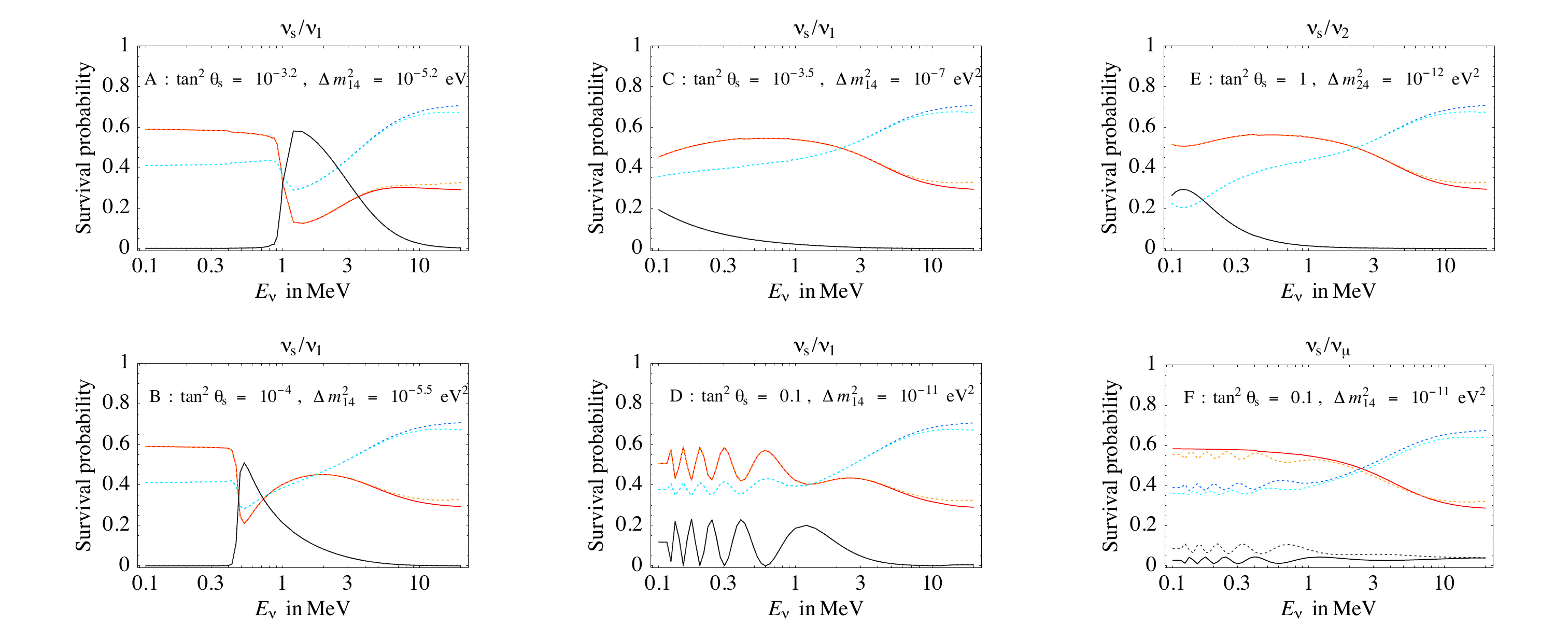}
    \caption{\footnotesize\em  Sample oscillations patterns of solar neutrinos with light 
    sterile neutrinos. 
    The decreasing red curves give $P(\nu_e\to\nu_e)$, 
    the increasing blue curves give  
$P(\nu_e\to\nu_{\mu,\tau}$) and the lower black curves give 
$P(\nu_e\to \nu_{\rm s})$.
The continuous (dotted) curve are the values during day (night).
From~\cite{cirelli}.   
     \label{fig:cir} }
  \end{figure}

Now, independently on whether we have better or worse theoretical motivations for light sterile neutrinos, it is possible to use solar neutrino data
to test whether we have any hint of {\em other} light neutrinos. A systematic analysis of this type was performed in 
\cite{cirelli}. Fig.~\ref{fig:cir}, taken from this reference, shows the  modification of the 
pattern of solar neutrino oscillations due to six possible sterile neutrinos, that were admitted by the data available in 2005. 
In view of the new data collected since then, it would be useful to repeat this type of analysis. 
This is interesting {\em per se} and potentially relevant in the context of global analyses aiming at finding hints of sterile neutrinos.

\section{Remarks}
The study of solar neutrinos is a peculiar branch of neutrino physics, with 
its own main achievements,  burning questions and dynamics of evolution.
It is linked with several disciplines such as experimental and theoretical particle physics, nuclear physics and astrophysics. 
These links contribute greatly to maintain solar neutrino physics lively.
As we have argued, solar neutrino physics is today 
in a healthy state,  just as  neutrino physics in general--see \cite{pol} for a wider discussion.

Coming to specific considerations, we have shown that a definitive understanding of how the Sun functions  
was obtained only 3 years ago by Borexino experiment at Gran Sasso laboratory. We mean here `understanding' in a  
Galilean sense, namely, we have been able to test the hypotheses (i.e., the theoretical expectations) by means of experiments and observations.
The same team is progressing with more goals ahead: We are on the verge of learning on CNO neutrinos. A lot of physics issues can be usefully investigated and surprises may still occur.
It should be not forgotten that such topics are of great interest for a very wide audience--it is easy to explain to a wide public that 
{\em we have understood how the Sun works}. Moreover, the Nobel prize in physics, assigned for the beginning of neutrino astronomy in 2002, 
witnesses that similar activities are also considered worth of reward by the scientific community.

The last remark is sociological in nature: 
Whether we want it or not, neutrino astronomy is largely in the hands of particle physicists.
This is true also, to a good extent, for the more specific branch of solar neutrino studies. Such a circumstance is not necessarily good or bad, however, in view of the healthy state of the field, it can be considered fortunate. It is also a reason of special responsibility: in particular, theorists working in particle physicists 
 are supposed to pay attention to the features of the field. In this manner, they will play an even more important role and will take great advantages from solar neutrinos.

\paragraph{Acknowledgments:} I thank G.~Barenboim,  
G.~G.~Raffelt for the invitation at CERN; 
A.~Strumia, F.~L.Villante for collaboration on solar neutrinos;
G.~Bellini, P.~Sapienza 
for precious discussions;
A.~Smirnov and L.~Lavoura for providing useful feedback.

\begin{multicols}{2}

\end{multicols}

\end{document}